# Pressure effect and Superconductivity in $\beta$-Bi$_4$I$_4$ Topological Insulator

A. Pisoni[1,*], R. Gaál[1], A. Zeugner[2], V. Falkowski[3], A. Isaeva[2], H. Huppertz[3], G. Autès[1], O. V. Yazyev[1] and L. Forró[1].

[1]*Institute of Physics, EPFL, CH-1015 Lausanne, Switzerland*

[2]*Department of Chemistry and Food Chemistry, TU Dresden, D-01062 Dresden, Germany*

[3]*Institute of General, Inorganic and Theoretical Chemistry, University of Innsbruck, A-6020 Innsbruck, Austria*

**Abstract**

We report a detailed study of the transport coefficients of $\beta$-Bi$_4$I$_4$ quasi-one dimensional topological insulator. Electrical resistivity, thermoelectric power, thermal conductivity and Hall coefficient measurements are consistent with the possible appearance of a charge density wave order at low temperatures. Both electrons and holes contribute to the conduction in $\beta$-Bi$_4$I$_4$ and the dominant type of charge carrier changes with temperature as a consequence of temperature-dependent carrier densities and mobilities. Measurements of resistivity and Seebeck coefficient under hydrostatic pressure up to 2 GPa show a shift of the charge density wave order to higher temperatures suggesting a strongly one-dimensional character at ambient pressure. Surprisingly, superconductivity is induced in $\beta$-Bi$_4$I$_4$ above 10 GPa with $T_c$ of 4.0 K which is slightly decreasing upon increasing the pressure up to 20 GPa. Chemical characterisation of the pressure-treated samples shows amorphization of $\beta$-Bi$_4$I$_4$ under pressure and rules out decomposition into Bi and BiI$_3$ at room-temperature conditions.

## I. INTRODUCTION

In the last years many bismuth binary compounds have been rediscovered as topological insulators (TIs) [1,2]. In TIs metallic surface states coexist with the bulk bandgap as a result of strong spin-orbit coupling. These surface states are topologically protected, which implies that they can be hardly altered by impurities or crystal defects, and exhibit unique transport phenomena [3]. We recently reported the discovery of a novel quasi-one dimensional (quasi-1D) topological insulator in $\beta$-Bi$_4$I$_4$ [4]. The crystal structure of $\beta$-Bi$_4$I$_4$ is composed of narrow one-dimensional metallic bismuth stripes that extend along the *b* crystallographic axis [5,6]. These 1D building blocks are held together by weak non-covalent bonds, forming 2D layers parallel to each other that in turn pile up making a 3D bulk single crystal [5,7]. Both first-principles electronic structure calculations and angle-resolved photoemission spectroscopy (ARPES) studies demonstrate that *β*-Bi$_4$I$_4$ is at the boundary of a strong-weak topological insulator phases and a trivial insulator one [4]. Pressure or chemical doping could drive the material in one or another phase inducing a topological phase transition [4,8] as already observed in *β*-As$_2$Te$_3$ [9] and PbTe [10]. Moreover, pressure can efficiently modify the small bandgap and the band structure of TIs, enhancing the thermoelectric properties, without introducing additional disorder like chemical doping would do [11].

---

[*] Corresponding author: andrea.pisoni@epfl.ch

In this work we report the temperature dependence of resistivity, thermoelectric power, thermal conductivity and Hall coefficient of *β*-Bi$_4$I$_4$ single crystals. Resistivity and thermoelectric power were measured as a function of temperature under hydrostatic pressure up to 2 GPa. Measurements of the electrical resistivity in quasi-hydrostatic pressure up to 20 GPa reveal that superconductivity emerges in *β*-Bi$_4$I$_4$ above 10 GPa. Chemical stability of pressure-treated *β*-Bi$_4$I$_4$ crystals was studied at various temperatures by means of X-ray diffraction and energy-dispersive X-ray analysis.

## II. EXPERIMENTAL DETAILS

Single crystals of *β*-Bi$_4$I$_4$ were obtained by chemical transport reaction employing bismuth metal and HgI$_2$ as starting materials as described in detail in refs. [4–6]. The samples appear with a needle-like shape with dimensions up to $10 \times 1 \times 0.5$ mm. High quality of the present crystals in terms of crystal structure and chemical composition was confirmed by the same palette of different characterisation methods as in [4]. Powder samples for combined pressure-temperature treatment were synthesized following [12].

The electrical resistivity ($\rho$) was measured as a function of temperature along the elongated direction (*b* axis) of *β*-Bi$_4$I$_4$ single crystals, in a conventional four point configuration using a delta-mode technique to eliminate thermoelectric voltages. Gold wires were attached to the samples by means of a graphite-based glue, in order to avoid any unwanted reaction between silver and iodine present in the samples. For the measurement of thermoelectric power ($S$) the sample was placed on an electrical insulating ceramic bar, with a small heater anchored at one end. The heat generated by the small resistor propagated through the sample along the *b* axis while the temperature gradient across it was measured by a differential type-E thermocouple, as described elsewhere [12]. To avoid any undesired reaction between the sample and the moisture in the air, the resistivity and thermoelectric power measurements, at ambient pressure, were performed in helium gas atmosphere. The thermal conductivity ($\kappa$) of $\beta$-Bi$_4$I$_4$ single crystals was measured by a steady-state method under vacuum (base pressure of 10$^{-6}$ mbar) along the *b* axis direction [13]. A small chip resistor was directly glued on one side of the sample using Stycast FT 2850. In order to measure the amount of heat passing through the specimen, a stainless steel reference sample of known thermal conductivity was connected between the sample and the copper sample holder, which acted as a heat sink. The temperature gradient on the reference sample and on the crystal was measured by type-E differential thermocouples. A temperature gradient of 1 K was carefully maintained across the sample at all temperatures. To measure the Hall coefficient ($R_H$) a thin single crystal of $\beta$-Bi$_4$I$_4$ was shaped into a standard six-terminal Hall bar configuration by focused ion beam (FIB) technique. Platinum leads were deposited on the sample by FIB. $R_H$ was measured applying a magnetic field of 1 T, perpendicular to the sample's *b* axis direction, that was reversed to eliminate the contribution from misaligned Hall contacts.

The temperature dependence of $\rho$ and *S* were measured simultaneously and on the same sample, in hydrostatic pressure up to 2 GPa provided by a piston-cylinder cell employing Daphne oil 7373 as pressure transmitting medium. The pressure was determined by the superconducting transition temperature of a lead pressure gauge. The electrical resistivity under high pressure was extended up to 20 GPa in a diamond anvil cell (DAC). For this scope electrical leads were implanted in the diamond culet by FIB and the sample was directly pressed on them during the measurement. Rhenium metal was used for the gasket, that was insulated from the platinum leads by a mixture of fine Al$_2$O$_3$ powder and black Stycast 2850 FT. NaCl powder was used as pressure transmitting medium. The pressure was measured at room temperature from the shift of the fluorescence line of a small ruby grain placed close to the sample [14].

After pressure release the sample with the transmission medium was extracted from the gasket hole and the diffraction data were taken from the rotating sample at ID23 ESRF beamline using 0.6888 Å wavelength at 173 K. PILATUS 6M detector was used; frames were binned by SNBL Tool Box software and azimuthal integration was done using Fit2D. Afterwards the chemical composition of the crystal was identified by energy dispersive X-ray analysis (EDX). The spectra were collected using Oxford Silicon Drift X-Max$^N$ detector at an acceleration voltage of 20 kV and a 100 s accumulation time. The EDX analysis was performed using the P/B-ZAF reference-free method (where Z = atomic no. correction factor, A = absorption correction factor, F = fluorescence factor, P/B = peak to background model). Scanning electron microscopy (SEM) was performed on a SU8020 (Hitachi) with a photodiode-low-energy-backscattered-electrons-detector to obtain evaluable Z-contrast ($U_a$ = 3 kV).

Additionally, a series of compression experiments on presynthesized $\beta$-Bi$_4$I$_4$ polycrystalline powders and selected single-crystals were performed in a multianvil device at different temperatures (from room temperature to 1273 K) and pressures (from 2 to 10 GPa) with different exposure times (from 30 to 270 minutes). For all experiments, Bi$_4$I$_4$ samples were filled into a molybdenum capsule (0.025 mm foil, 99.95%, Alfa Aesar, Karlsruhe, Germany). The capsule was transferred into a crucible made from hexagonal boron nitride (HeBoSint® P100, Henze BNP GmbH, Kempten, Germany), built into an 18/11 assembly, and compressed by eight tungsten carbide cubes (HA-7%Co, Hawedia, Marklkofen, Germany). Further information on the construction of the assembly can be found in reference [15–18]. To apply pressure, a hydraulic press (mavo press LPR 1000-400/50, Max Voggenreiter GmbH, Mainleus, Germany) and a modified Walker-type module (also Max Voggenreiter GmbH) were used. Depending on the requested pressure, compression and decompression took place from 60 to 240 and 180 to 700 minutes, respectively. Additional heating was performed during the interjacent exposure time. Crystallinity and phase composition of these samples were studied by X-ray powder diffraction (Stoe Stadi P powder diffractometer with Ge(111) monochromatized Mo$K_{\alpha 1}$ ($\lambda$ = 70. 93 pm) radiation equipped with a Mythen 1K detector).

### III. RESULTS AND DISCUSSION

#### A. Ambient pressure transport properties

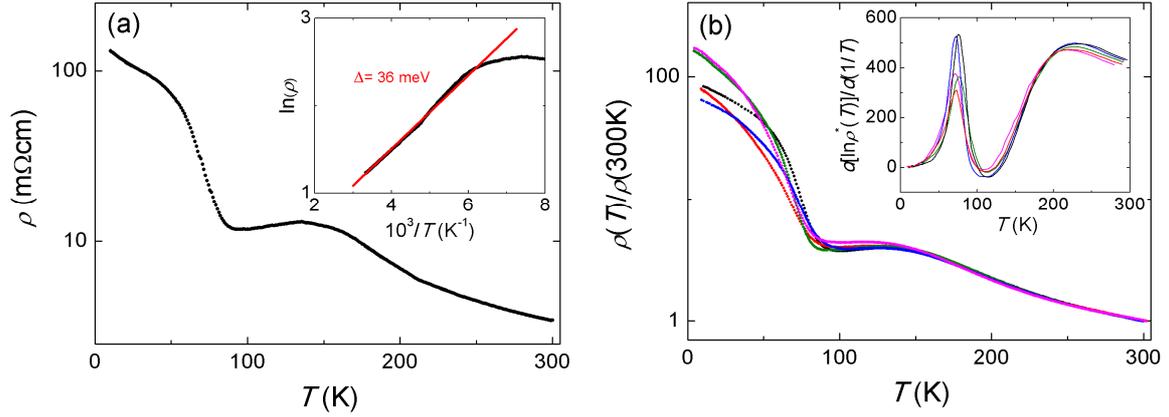

FIG. 1. (Color online) (a) Temperature-dependent resistivity of a β-Bi4I4 single crystal. The inset shows the logarithm of the resistivity as a function on the inverse of temperature. The fitting of the curve with a thermally activated model (red line) gives an activation energy (Δ) of 36 meV. (b) Temperature dependence of the resistivity of other five β-Bi4I4 single crystals normalized to the value at 300 K in order to emphasize high temperature region. The different curves overlap very well, indicating a same value of thermal activation energy, as described in the text. The inset presents a comparison between the derivatives of the logarithm of the different renormalized resistivity curves as a function of the temperature. The position of the peak attributed to the suspected CDW transition ($T_{CDW}^{\rho}$) doesn't change appreciably between different samples.

Fig. 1(a) presents the electrical resistivity of β-Bi4I4 measured in the 4–300 K temperature range, in logarithmic scale. Three different regimes can be identified: for temperatures above 150 K β-Bi4I4 displays a semiconducting behaviour ($d\rho/dT < 0$) while at lower temperature $\rho(T)$ starts decreasing smoothly down to 80 K where a sharp insulating transition increases its value by one order of magnitude. As $T$ further decreases another mild change of slope appears around 50 K. The fitting of $\rho(T)$ for $T > 150$ K with a thermally activated model $\rho(T) = \rho_0 \exp(\Delta/k_B T)$ (where $\Delta$ is the activation energy and $k_B$ the Boltzmann constant), gives a value of $\Delta = 36$ meV (inset of Fig. 1(a)), as we already reported [4]. This value is astonishingly close to that of the small direct bandgap at the Y point of the bulk Brillouin zone estimated by ab-initio calculations and confirmed by ARPES measurements [4]. In order to check the repeatability of our result we measured $\rho(T)$ of other five β-Bi4I4 single crystals having different dimensions. The results are presented in different colours in Fig. 1(b). The resistivity curves are normalized to their value at 300 K to emphasize the high temperature region. As it can be noted, the different curves perfectly overlap at high temperature indicating the same value of $\Delta$. Filatova *et al.* [5] suggested that the steep increase of $\rho(T)$ around 80 K could be caused by a charge-density-wave (CDW) transition in β-Bi4I4: in fact the quasi-1D character of this compound would facilitate the appearance of such an ordered state. In order to investigate the suspected CDW transition we plotted the derivatives of the logarithm of the different renormalized resistivity curves, $d(\ln\rho^*)/d(1/T)$ (where $\rho^* = \rho(T)/\rho(300\text{K})$, as a function of $T$ (inset of Fig. 1(b)). A sharp peak around $T^{\rho}_{CDW} \sim 80$ K can be clearly identified in the derivative of the different curves, very similarly to what observed in other CDW materials [19]. The value of $T^{\rho}_{CDW}$ does not appear to change significantly between. However, the resistivity flattening at lower temperature is more or less pronounced and occurs at different temperatures in different samples, suggesting that the presence of shallow in-gap impurity levels could be the cause of it.

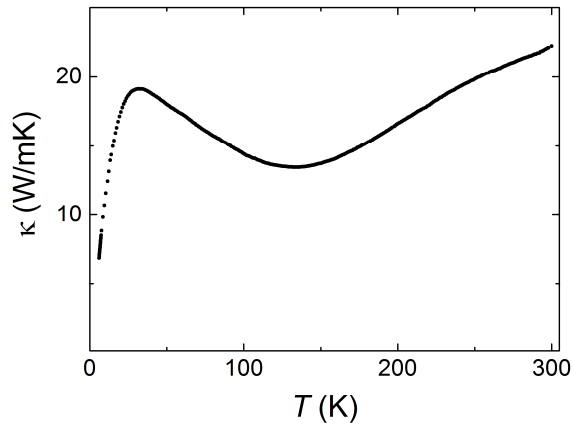

FIG. 2. (Color online) Temperature dependence of the thermal conductivity of a $\beta$-Bi$_4$I$_4$ single crystal. The broad minimum supports the presence of a CDW order.

The thermal conductivity of $\beta$-Bi$_4$I$_4$ (Fig. 2) has a value of $\kappa(300\text{K}) = 22$ WK$^{-1}$m$^{-1}$, that is very high for a quasi-1D material and it is even higher than that of bulk Bi$_2$Te$_3$ ($\sim 2$ WK$^{-1}$m$^{-1}$ [20]) and BiTeCl ($\sim 3$ WK$^{-1}$m$^{-1}$ [21]). At high temperature $\kappa$ decreases almost linearly down to a broad minimum around 135 K, below which it recovers the temperature dependence predicted by the Callaway theory [13,22]. On general basis, the total $\kappa$ is composed of an electrical and a phononic contribution. As for other poorly conducting materials, we can expect that the dominant contribution to $\kappa$ is given by phonons [19]. Indeed, assuming an electron density that justifies the usage of the Wiedemann–Franz law, one obtains a very low estimation for the value of the thermal conductivity due to charge carriers, being much less than 1 WK$^{-1}$m$^{-1}$. Therefore, one can safely neglect this term and analyse the $\kappa$ behaviour in terms of purely phonon contribution. The maximum that appears around 32 K is a consequence of Umklapp phonon–phonon scattering, which reduces the phonon mean free path at high temperature where the phonon thermal occupation is high for almost all wavelengths. By lowering $T$, the phonon thermal occupation decreases and Umklapp scattering becomes less significant, which results in an overall rise of thermal conductivity. At very low temperature, only the long-wave phonons remain in the system, with a mean free path that is constrained by the sample dimensions. The maximum in $\kappa$ at low temperature is experimentally found to occur around 10% of the Debye temperature ($\theta_D$) [19]. Therefore we can estimate a $\theta_D \sim 320$ K for $\beta$-Bi$_4$I$_4$.

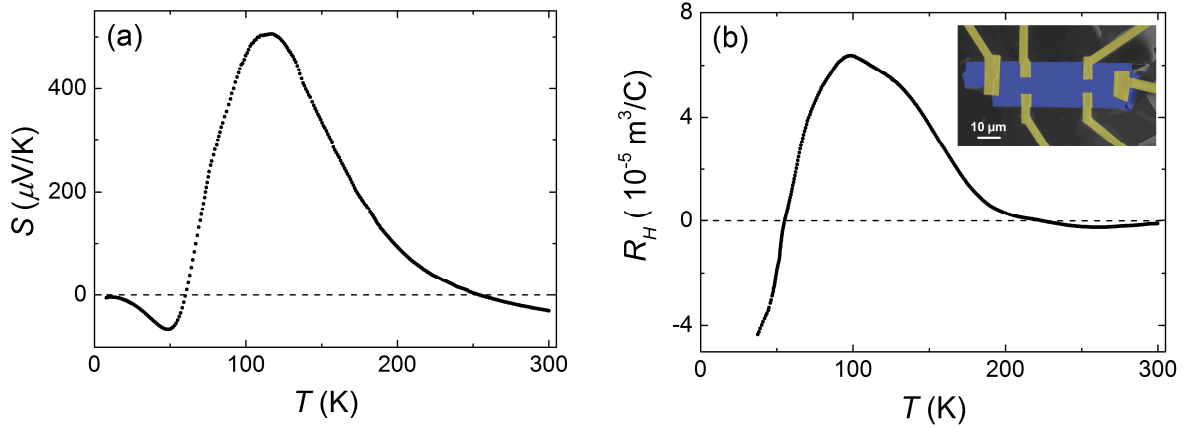

FIG. 3. (Color online) (a) Thermoelectric power of $\beta$-Bi$_4$I$_4$ single crystal as a function of temperature. (b) Temperature dependent Hall coefficient of $\beta$-Bi$_4$I$_4$ single crystal. The inset shows a SEM image of the sample contacted by FIB in a Hall bar configuration. The sample is highlighted in blue and the deposited electrical contacts in yellow.

The striking feature of $\kappa$ is the linear decrease at high $T$ which is quite unusual for a highly crystalline sample. It could be considered as the sign of the presence of CDW fluctuations which diffuse the phonons. At the establishment of a long range CDW order this source of scattering disappears and $\kappa$ follows the $T$ dependence expected for crystalline solids. It has to be mentioned that the $T_{CDW}^{\rho}$ is identified to be around at 80 K, but $\kappa$ starts to increase slightly already at 100 K. It means that the building up interchain CDW correlations is already sufficient to decrease the phonon scattering. Very similar observations were reported in other CDW compounds like 2H-TaSe$_2$ [23], (NbSe$_4$)$_3$I and K$_{0.3}$MoO$_3$ [24]. The significant difference between the CDW transition temperature observed in $\rho(T)$ and in $\kappa(T)$ can be due to fluctuation effects. Specific heat measurements on K$_{0.3}$MoO$_3$ demonstrated that fluctuation effects can appear above $T_{CDW}$ over a temperature interval that can be as much as 30 K wide [25].

Both the negative Seebeck (Fig. 3(a)) and the Hall (Fig. 3(b)) coefficients confirm that electrons are the dominant type of charge carriers in β-Bi$_4$I$_4$ at high temperature. As $T$ decreases, $S$ and $R_H$ increase becoming positive below 260 K and 230 K, respectively. This indicates that, indeed, both electrons and holes contribute to the electrical conduction in β-Bi$_4$I$_4$. The thermoelectric power and the Hall coefficient for a nondegenerate semiconductor in a two-band model, consisting of electron and hole bands, can be expressed as [26,27]:

$$S(T) = -\frac{k_B}{e}\left(\frac{\mu_e - \mu_h}{\mu_e + \mu_h}\right)\left(\frac{\Theta}{k_B T} + \text{constant}\right), \tag{1.1}$$

and

$$R_H = \frac{1}{e}\frac{n_h \mu_h^2 - n_e \mu_e^2}{(n_h \mu_h + n_e \mu_e)} \tag{1.2}$$

where $e$ is the electron charge, $\mu_e(\mu_h)$ is the electron (hole) mobility, $n_e(n_h)$ the electron (hole) density and Θ is the activation energy (in principle different from $\Delta$). Equations 1.1 and 1.2 can successfully describe the temperature dependence of $S$ and $R_H$ as the balance between the hole and electron bands is changed by $T$-dependent mobilities and carrier density if the charges respond differently to phonons or charge fluctuations. Unfortunately, without any additional knowledge, there is no mean to determine $\mu_e(\mu_h)$ and $n_e(n_h)$. The fluctuation effects associated with the CDW represent an extra source of entropy that could be the cause of the strong enhancement of $S(T)$ and $R_H(T)$ of β-Bi$_4$I$_4$ around $T_{CDW}^{\rho}$ [28]. Around 60 K the thermoelectric power and the Hall coefficient turn negative again. At this same temperature $\rho(T)$ shows a change of slope. This further supports our hypothesis that the flattening of $\rho(T)$ at low temperature is caused by shallow donor level just below the conduction band edge. Knowing the transport coefficients, $\rho$, $S$ and κ one can calculate the dimensionless thermoelectric figure of merit, $ZT = \frac{S^2}{\rho \kappa}T$. In order to have a promising material for thermoelectric applications values of $ZT \geq 1$ are necessary [29]. β-Bi$_4$I$_4$ presents $ZT(300\text{K}) = 4\times10^{-4}$ that increases only up to $1.8\times10^{-2}$ at 125 K where a maximum appears. Below this temperature $ZT$ further decreases. These values mean that β-Bi$_4$I$_4$ is not a promising thermoelectric material [29] which is not surprising because of the high values of thermal conductivity and electrical resistivity.

## B. Pressure effect

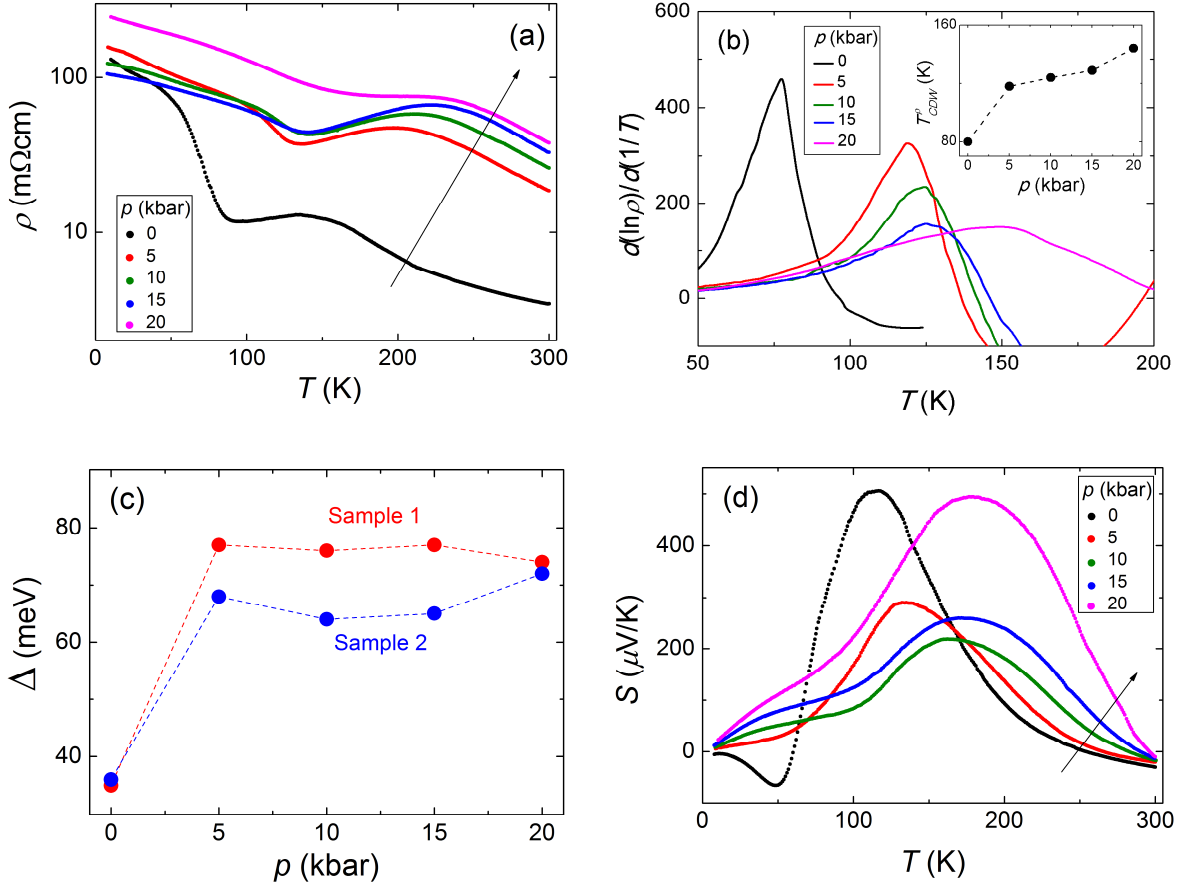

FIG. 4. (Color online) (a) Electrical resistivity of β-Bi$_4$I$_4$ as a function of temperature, measured at different pressures. The pressure is increased following the black arrow. (b) The derivative of the logarithmic electrical resistivity of β-Bi$_4$I$_4$ measured at different pressures and plotted against $T$, at low temperatures. The temperature at which the peak occurs in the different curves marks the CDW phase transition temperature $T^\rho_{CDW}$. The inset presents the pressure dependence of $T^\rho_{CDW}$, the black dashed line is a guide to the eyes. (c) Pressure dependence of the thermal activation energy extracted by resistivity measurements for two different samples. The results appear quite reproducible. Coloured dashed lines are guides to the eyes. (d) Temperature dependence of the thermoelectric power of β-Bi$_4$I$_4$ measured at different pressures. The pressure is increased following the black arrow.

The temperature dependence of the electrical resistivity measured in hydrostatic pressure up to 2 GPa, is shown in Fig. 4(a). Hydrostatic pressure ($p$) monotonously increases the resistivity at room temperature from $\rho(300\,\text{K}) = 3\,\text{m}\Omega\text{mm}$ at ambient pressure to $38\,\text{m}\Omega\text{mm}$ at 2 GPa and shifts the local maximum in $\rho$ to higher temperatures. In order to study the effect of pressure on the CDW transition we plotted $d(\ln\rho)/d(1/T)$ as a function of $T$ for different pressures (Fig. 4(b)). The CDW transition temperatures at different pressures can be clearly identified as the temperatures where the peak in the derivative occurs. The inset of Fig. 4(b) presents the pressure dependence of $T^{\rho}_{CDW}$. With increasing pressure the height of the peaks in Fig. 4(b) is greatly reduced. This would indicate that pressure weakens the CDW transition in *β*-Bi$_4$I$_4$, in very good agreement with the general trend reported for other CDW materials [30,31]. At first sight, the monotonous increase of $T^{\rho}_{CDW}$ with pressure seems to contradict the previous conclusion. A similar enhancement of the CDW transition temperature by pressure was already observed in (NbSe$_4$)$_{10/3}$I [32], (TaSe$_4$)$_2$I [33] and in some organic charge-transfer salts [34]. A tentative explanation of this observation is the following: in quasi-1D systems $T^{\rho}_{CDW}$ is a function of the dimensionality, that is of the interchain coupling [35], and pressure can cause an increase of $T^{\rho}_{CDW}$ by suppressing the 1D fluctuations both directly or indirectly [33]. In the ultimate 1D system there would be no CDW ordering because large fluctuations preclude all long range order, while in 3D, by definition, there is no 1D phase transition. In consequence, depending on the interchain coupling (degree of one-dimensionality) there should be a maximum in $T^{\rho}_{CDW}(p)$. From our measurements it seems that our system is on the one-dimensional side of $T^{\rho}_{CDW}(p)$, and pressure is bringing it to the maximum value of $T^{\rho}_{CDW}$ but without reaching it. Fig. 4(c) shows the evolution of the thermal activation energy, extracted from $\rho$ at high temperature, as a function of pressure for two different samples. The measurement of $\Delta$ under pressure is particularly important because first-principles calculations [4] predict a decrease of $\Delta$ above a critical pressure value, as evidence of a possible weak–strong topological phase transition. However, in the 0–2 GPa pressure range we do not observe it.

The temperature dependence of the Seebeck coefficient is already very complex at ambient pressure and it remains as such at different pressures (Fig. 3(b)). We suppose that it is the change of the subtle balance in the contribution of the two types of carriers in temperature and pressure. Pressure induces only a small change in $S(300\text{K})$ that becomes less negative. The main changes are visible in the temperature $\left(T_{max}^{S}\right)$ and the magnitude $\left(S_{max}\right)$ at the maximum of $S$. Pressure moves $T_{max}^{S}$ to higher temperature in the same way as $T_{CDW}^{\rho}$. This confirms that the maximum of $S$ is somehow related to the appearance of the CDW state. The value of $S_{max}$ first decreases from 505 µV/K at 0 kbar to 218 µV/K at 10 kbar and for higher pressures it increases again reaching a value of 495 µV/K at 20 kbar. The abrupt increase of $S_{max}$ at 20 kbar could be associated with the less conducting behaviour of $\rho(T)$ at the same pressure. Below 50 K the Seebeck coefficient changes sign passing from negative to positive above 5 kbar. As *p* further increases $S$ remains positive and the slope at which it approaches zero increases. This indicates that in *β*-Bi4I4 for *p*> 5 kbar the holes become the dominant charge carriers at low-*T*.

### C. Pressure-induced superconductivity

Superconductivity was induced in *β*-Bi4I4 under quasi-hydrostatic pressure provided by a diamond anvil cell (DAC). The temperature-dependent resistance of a *β*-Bi4I4 single crystal at different pressures up to 20 GPa is shown in Fig. 5(a). The overall behaviour recalls that recently published by Zhou *et al.* [36] about ZrTe5 three-dimensional topological insulator and by various groups [37,38] for the Rashba-material BiTeI that undergoes a topological transition under pressure. The *β*-Bi4I4 sample was mechanically pressed on the electrical contacts and for this reason it was not possible to measure at pressures lower than 5 GPa. Already at 5 GPa the resistance has a *T*-dependence very different from that seen at 2 GPa in Fig. 4(a). At high-*T* some reminiscent insulating behaviour can still be noticed and $R(T)$ decreases its slope around 150 K. For $p \geq 8\,\text{GPa}$, $R(T)$ presents an almost temperature-independent behaviour. The value of $R(300\,\text{K})$ does not follow a monotonous trend: it first increases up to $p = 10\,\text{GPa}$ and then it non-continuously decreases for $p > 10\,\text{GPa}$. These two observations, especially the almost temperature-independent resistivity, which would mean a high level of static disorder, make us think that the sample lost its crystallinity with pressure.

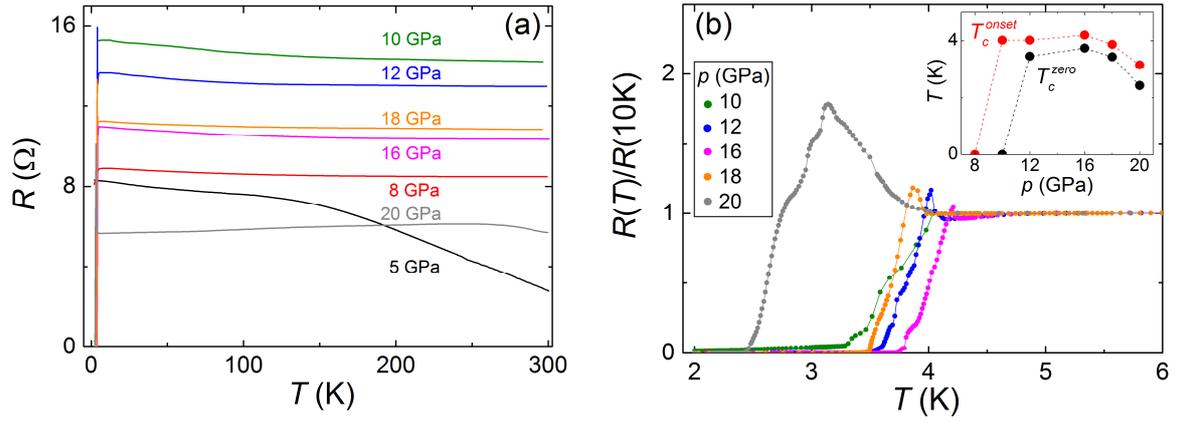

FIG. 5. (Color online) (a) Temperature dependence of the resistance of a $\beta$-Bi$_4$I$_4$ single crystal measured at high pressure. (b) Detailed view of the resistance curves normalized to their value at 10 K to emphasize the change in $T_c$ as a function of pressure. The inset shows the pressure dependence of the onset and zero-resistance critical temperatures.

This has fingerprints in the appearance of superconductivity with pressure, as well. Figure 5(b) shows the resistance curves normalized to their value at 10 K to emphasize the change in the critical temperature ($T_c$) as a function of pressure. A superconductivity onset is observed around 4 K at 10 GPa. In the following analysis we define $T_c^{onset}$ as the temperature extracted from the intersection of two linear fits: one drawn in the normal state just above the superconducting transition and one in the steepest part of the transition. For $p > 10\,\text{GPa}$ we identify $T_c^{onset}$ with the temperature of the peak appearing just above the superconducting transition. Moreover we define $T_c^{zero}$ as the temperature at which $R(T) \sim 0$, namely when $R(T)$ falls below the sensitivity of our instruments. At 10 GPa, $R(T)$ doesn't approach zero, presumably indicating a filamentary type of superconductivity. At 12 GPa the zero-resistance state sets in and persists up to 20 GPa. Already at 12 GPa we observe a sharp peak appearing just above the transition which is the fingerprint of the non-uniformity of the superconductivity. Furthermore, although the superconducting transition width is always less than 1 K for all applied pressures, some steps can be seen in the drop of $R(T)$. Inhomogeneity of the pressure distribution over the sample could be the cause of it. The inset in Fig. 4(b) shows the pressure dependence of $T_c^{onset}$ and $T_c^{zero}$. The T–p phase diagram presents a nice dome-like structure suggesting an optimum $T_c^{onset} \sim 4.2$

K at 16 GPa, that decreases to 3.2 K at 20 GPa. We were able to tune $T_c$ of $β$-Bi$_4$I$_4$ reversibly by increasing and decreasing pressure up to 20 GPa.

### D. Chemical characterization of the pressure-treated samples

The observed $T$-independent resistivity and the inhomogeneities seen in the superconducting transition predicated a study of pressure-dependent chemical stability of $β$-Bi$_4$I$_4$. Three possible scenarios of chemical alternation were taken into account: amorphisation of $β$-Bi$_4$I$_4$, structural transition between $β$-Bi$_4$I$_4$ and $α$-Bi$_4$I$_4$ modifications of the bismuth monoiodide, and decomposition into elemental bismuth and BiI$_3$ which occurs at ca. 603 K [1] at ambient pressure. Particularly, the last instance would be discouraging since amorphous bismuth precipitates could then support superconductivity [1,39]. The critical temperature $T_c$ of the elemental bismuth constitutes 7 K at 5 GPa and slightly decreases upon increasing pressure [34]. Hence the discrepancy observed in our experiments (4 K vs. 7 K) could be in principle ascribed to a size effect of Bi particles. BiI$_3$ which would be the other decomposition product is fairly air-sensitive and would lead to rapid hydrolysis and oxidation of the sample in air. The resistivity of BiI$_3$ under pressures up to 3.3 GPa was recently studied elsewhere [1].

First and foremost, the sample treated at 20 GPa was analysed via a synchrotron X-ray diffraction measurement performed at the European Synchrotron Radiation Facility (ESRF). Regardless obvious physical integrity of the crystal and its smooth surface without signs of deterioration (Fig. 6(a)), all observed Bragg reflections could be assigned to ruby (Al$_2$O$_3$) and NaCl, thus hinting at amorphization of bismuth monoiodide under pressure (Fig. 6(b)). Nevertheless, two consecutive EDX analyses of the sample (with a time interval of 7 months) confirmed very uniform distribution of bismuth and iodine with the ratio close to 1:1 in both measurements (Fig. 7). The surface appears moderately oxidized, thus ruling out any significant admixtures of BiI$_3$. Presence of any Bi crystallites was not observed. These results support that at least chemical composition of Bi$_4$I$_4$ is preserved and thus an amorphous phase of Bi$_4$I$_4$ could support superconductivity, for instance, like in some amorphous actinide alloys [35]. This possibility is very interesting by itself, and it could be the subject of an upcoming study.

(b)

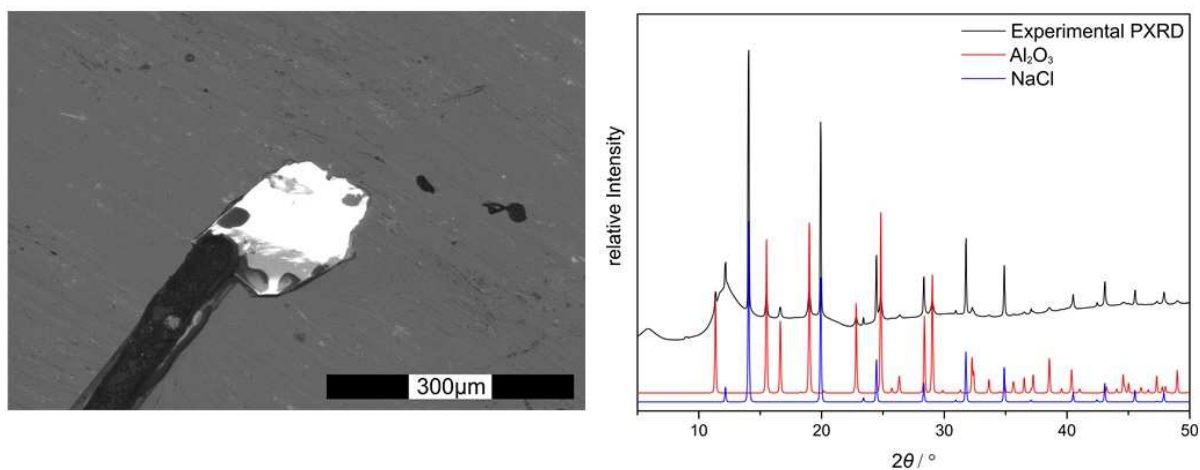

FIG. 6. (Color online) (a) Bismuth monoiodide crystal after the pressure treatment at 20 GPa (SEM micrograph, 3kV, PD-BSE detector) (b) Corresponding synchrotron X-ray diffraction pattern.

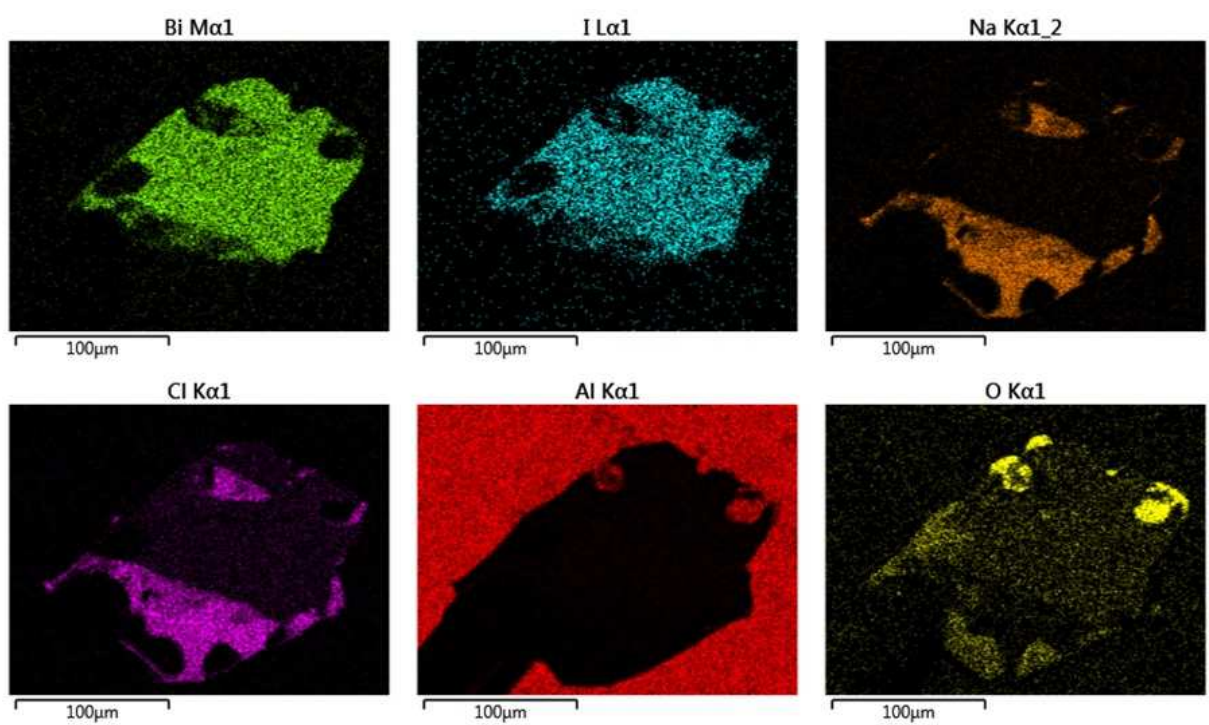

FIG. 7. (Color online) EDX mapping spectra for the $β$-$Bi_4I_4$ crystal after high-pressure treatment (first measurement; see text). Note distinct areas corresponding to admixtures of NaCl and $Al_2O_3$. The accuracy of the composition determination was enhanced thanks to NaCl acting as an internal reference. Averaged results of the first measurement: Bi 47.4 at. %, I 52.6 at. %; Na 53.0 at. %, Cl 47.0 at. %. Averaged results of the second measurement (performed after 7 months): Bi 48.0 at. %, I 52.0 at. %; Na 53.3 at. %, Cl 46.7 at. %.

Furthermore, pressure and temperature-induced effects on structure and composition of the starting material $β$-$Bi_4I_4$ were studied on larger amounts of presynthesized well-crystalline samples using the multianvil technique. These experiments are summarized in Table I. Experiments at room temperature and 7.5 GPa with different exposure times (30 and 180 min, respectively) reveal time-dependent amorphisation of the sample (see the results of X-ray powder diffraction measurements in Fig. 8). The reflection intensities drop significantly with increasing exposure time. No indication for formation of other decomposition products was found. As no in situ measurements were carried out during these experiments, it was not possible to make a statement, whether pressure-induced amorphisation of $Bi_4I_4$ took place or if an amorphous decomposition product of an intermediate phase, which is unstable at ambient pressure, had formed.

TABLE I. Conditions for pressure experiments and identified products.

| Pressure (GPa) | Temperature (K) | Compression/ decompression time (min) | Exposure time (min) | Identified phases |
|---|---|---|---|---|
| | | Single crystals | | |
| 3 | 473 | 70/210 | 70 | $\alpha$-, $\beta$-$Bi_4I_4$ |
| 3 | 623 | 70/210 | 70 | $\alpha$-, $\beta$-$Bi_4I_4$, Bi, $BiI_3$ |
| 5 | 1023 | 130/365 | 270 | Bi, $BiI_3$ |
| 7.5 | 293 | 210/600 | 30 | $\alpha$-, $\beta$-$Bi_4I_4$ |
| 7.5 | 293 | 210/600 | 180 | $\alpha$-, $\beta$-$Bi_4I_4$ |
| 10 | 443 | 235/700 | 70 | Bi, $BiI_3$ |
| 10 | 553 | 235/700 | 85 | Bi, $BiI_3$ |
| | | Powdered samples | | |
| 2 | 873 | 60/180 | 80 | Bi, $BiI_3$ |
| 3 | 453 | 70/210 | 65 | $\alpha$-, $\beta$-$Bi_4I_4$ |
| 3 | 573 | 70/210 | 65 | $\alpha$-, $\beta$-$Bi_4I_4$ |
| 5 | 573 | 130/365 | 120 | $\alpha$-, $\beta$-$Bi_4I_4$, Bi, $BiI_3$ |
| 5 | 673 | 130/365 | 90 | Bi, $BiI_3$ |
| 5 | 1023 | 130/365 | 210 | Bi, $BiI_3$ |
| 5 | 1273 | 130/365 | 65 | Bi, $BiI_3$ |
| 7.5 | 523 | 210/600 | 100 | Bi, $BiI_3$ |

Further multianvil experiments at different pressures (2, 3, 5, 7.5, and 10 GPa) and an energy input in form of heating (from 443 to 1273 K) revealed a decomposition of $Bi_4I_4$ into Bi and $BiI_3$ at 7.5 GPa and above, whereas between 2 and 5 GPa, a mixture of β-$Bi_4I_4$ and α-$Bi_4I_4$ survived the annealing at temperatures below ca. 623 K showing a strong tendency towards a rapid amorphisation.

Also at higher temperatures ($T$ = 673 – 1273 K), decomposition in Bi and $BiI_3$ took place in the pressure range of 2–5 GPa. Furthermore, $BiI_3$ needle-shaped crystals could be grown at all above mentioned pressures and temperatures between 973 and 1273 K.

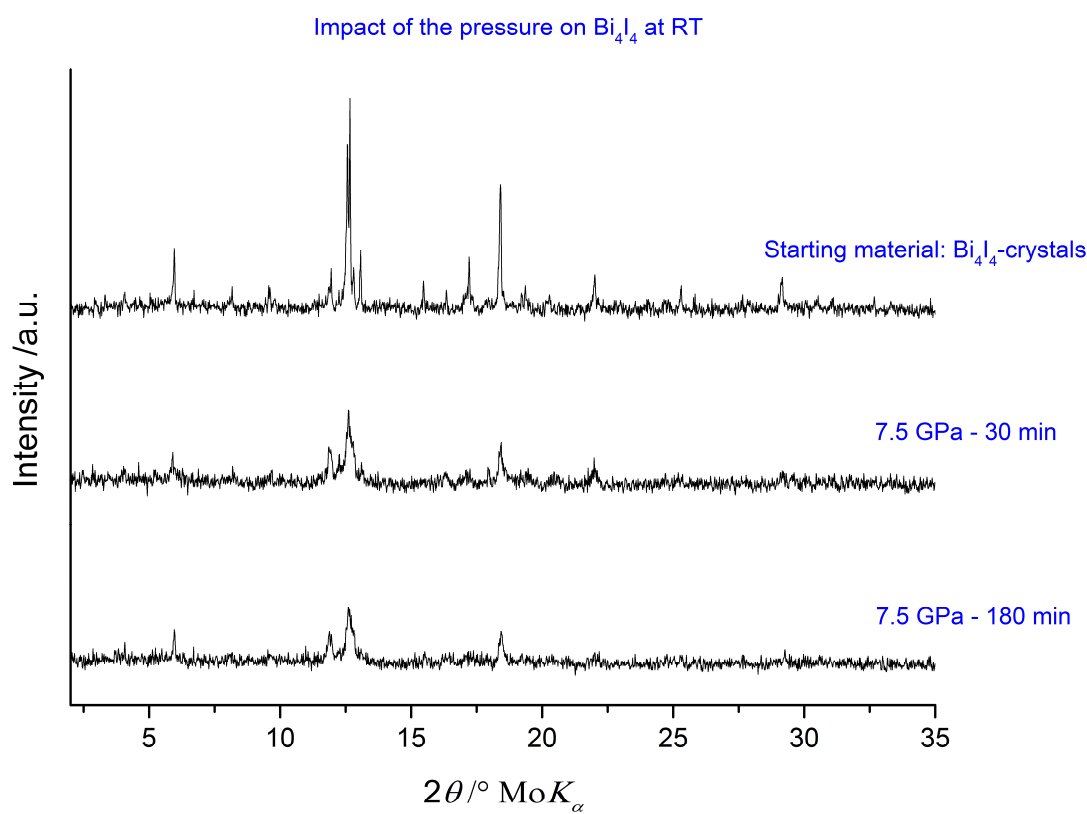

FIG. 8. (Color online) PXRD study of the impact of high-pressure treatment on β-$Bi_4I_4$ crystals at room temperature.

## IV. CONCLUSIONS

We investigated the transport coefficients of the quasi-one dimensional *β*-$Bi_4I_4$ single crystals. Electrical resistivity, thermoelectric power, thermal conductivity and Hall coefficient reveal a very rich temperature dependence which is tentatively ascribed to CDW formation. Seebeck and Hall coefficients show that both electrons and holes contribute to the conduction in *β*-$Bi_4I_4$ and that the dominant type of charge carriers changes with temperature as a consequence of changes in the mobilities and densities of the two types of carriers. Application of hydrostatic pressure up to 2 GPa shifts the CDW transition in *β*-$Bi_4I_4$ to higher temperatures, as demonstrated by resistivity and thermoelectric power measurements. Resistivity measurements up to 20 GPa reveal that superconductivity is induced in *β*-$Bi_4I_4$ at 10 GPa although the material loses its crystallinity and retains its composition. A maximum critical temperature of 4.2 K is reached at 16 GPa, which decreases slightly upon further increase of pressure. Decomposition of *β*-$Bi_4I_4$ into Bi and $BiI_3$ occurs in the 2–7.5 GPa range only at elevated temperatures. Despite this detailed study of the transport coefficients we could not pinpoint a particular property which would be typical for the topological insulating nature of *β*-$Bi_4I_4$ reported previously.

## ACKNOWLEDGEMENT

This work was supported by the Swiss National Science Foundation (Project No. 200021_138053) and by the German Research Foundation (DFG) in the framework of the Special Priority Programme (SPP 1666) "Topological Insulators" and by the ERA-Chemistry Programme (DFG project number RU 776/15-1 and FWF project number I 2179) . We are grateful to Dr. Alexey Bosak (ESRF, Grenoble, France) for the synchrotron measurements of our sample and to Mr. Alexander Weiz and Mrs. Michaela Munch (TU Dresden, Germany) for synthetic help and valuable discussions.